\documentclass[useAMS,usenatbib]{JHEP3}
\usepackage{times}
\usepackage{graphicx}
\usepackage{subfigure}
\usepackage{psfrag}
\usepackage{amsmath}
\usepackage{amsfonts}   
\usepackage{amssymb}    
\usepackage[numbers ,square, comma, sort&compress]{natbib} 

\def\reff@jnl#1{{\rm#1\/}}
\def\aj{\reff@jnl{AJ}}                 
\def\araa{\reff@jnl{ARA\&A}}           
\def\apj{\reff@jnl{ApJ}}               
\def\apjl{\reff@jnl{ApJ}}              
\def\apjs{\reff@jnl{ApJS}}             
\def\ao{\reff@jnl{Appl.Optics}}        
\def\apss{\reff@jnl{Ap\&SS}}           
\def\aap{\reff@jnl{A\&A}}              
\def\aapr{\reff@jnl{A\&A~Rev.}}        
\def\aaps{\reff@jnl{A\&AS}}            
\def\azh{\reff@jnl{AZh}}               
\def\baas{\reff@jnl{BAAS}}             
\def\jrasc{\reff@jnl{JRASC}}           
\def\memras{\reff@jnl{MmRAS}}          
\def\mnras{\reff@jnl{MNRAS}}           
\def\pra{\reff@jnl{Phys.Rev.A}}        
\def\prb{\reff@jnl{Phys.Rev.B}}        
\def\prc{\reff@jnl{Phys.Rev.C}}        
\def\prd{\reff@jnl{Phys.Rev.D}}        
\def\prl{\reff@jnl{Phys.Rev.Lett}}     
\def\pasp{\reff@jnl{PASP}}             
\def\pasj{\reff@jnl{PASJ}}             
\def\qjras{\reff@jnl{QJRAS}}           
\def\skytel{\reff@jnl{S\&T}}           
\def\solphys{\reff@jnl{Solar~Phys.}}   
\def\sovast{\reff@jnl{Soviet~Ast.}}    
\def\ssr{\reff@jnl{Space~Sci.Rev.}}    
\def\zap{\reff@jnl{ZAp}}               
\def\nat{\reff@jnl{Nature}}            

\renewcommand{\Pr}{P}

\title{Challenges of Profile Likelihood Evaluation in Multi--Dimensional SUSY Scans}
\author{Farhan Feroz\\
	  Cavendish Laboratory, JJ Thomson Avenue, Cambridge CB3 0HE, UK\\
        E-mail: \email{f.feroz@mrao.cam.ac.uk}}  
\author{Kyle Cranmer\\
	  Center for Cosmology and Particle Physics, New York University, Washington Place, New York, NY 10003, USA\\
        E-mail: \email{cranmer@cern.ch}}
\author{Mike Hobson\\
	  Cavendish Laboratory, JJ Thomson Avenue, Cambridge CB3 0HE, UK\\
        E-mail: \email{mph@mrao.cam.ac.uk}}
\author{Roberto Ruiz de Austri\\
	  Instituto de F\'{\i}sica Corpuscular, IFIC-UV/CSIC, Valencia, Spain\\
        E-mail: \email{rruiz@ific.uv.es}}
\author{Roberto Trotta\\
	  Imperial College London, Blackett Laboratory, Prince Consort Road, London, SW7 2AZ, UK\\
        E-mail: \email{r.trotta@imperial.ac.uk}}

\abstract{Statistical inference of the fundamental parameters of supersymmetric theories is a challenging and active endeavor. 
  Several sophisticated algorithms have been employed to this end.  While Markov-Chain Monte Carlo (MCMC) and nested sampling
  techniques are geared towards Bayesian inference, they have also been used to estimate frequentist confidence intervals based
  on the profile likelihood ratio.
  We investigate the performance and appropriate configuration of {\sc
  MultiNest}, a nested sampling based algorithm, when used for profile likelihood-based analyses both on toy models and on the
  parameter space of the Constrained MSSM.  We find that while the standard configuration previously used in the literarture is appropriate for an accurate
  reconstruction of the Bayesian posterior, the profile likelihood is poorly approximated.  We identify a more appropriate {\sc
  MultiNest} configuration for profile likelihood analyses, which gives an excellent exploration of the profile likelihood
  (albeit at a larger computational cost), including the identification of the global maximum likelihood value. We conclude that
  with the appropriate configuration {\sc MultiNest} is a suitable tool for profile likelihood studies, indicating previous
  claims to the contrary are not well founded.}


\begin{document}
\label{firstpage}

\section{Introduction}\label{sec:intro}

The results of the first searches for supersymmetry (SUSY) at the Large Hadron Collider (LHC) have recently been
released~\cite{Collaboration:2011tk}.  The field is eager to switch from a mode of excluding model parameters to an exciting and challenging era in
which we are estimating the underlying parameters of a new fundamental theory.  Within the context of SUSY, the technology has evolved from likelihood
scans over low-dimensional subspaces of SUSY models~\cite{Roberts:1993tx,Kane:1993td,Ellis:2003si,Profumo:2004at} to higher-dimensional Bayesian
analysis using Markov Chain Monte Carlo (MCMC)~\cite{2004JHEP...10..052B,Allanach:2005kz,2006JHEP...05..002R}.  More recently, the nested sampling
\cite{skilling04, Sivia} algorithm has been used by several authors for the study of SUSY models~\cite{2009PhRvD..80c5017A, 2010PhRvD..81i5012A,
2009arXiv0903.2487F, 2008JHEP...12..024T, 2008JHEP...10..064F, 2010JHEP...07..064W, 2010arXiv1010.2023B, 2010PhRvD..82e5003R, 2010JCAP...01..031S,
2010arXiv1012.3939R}.  Although Bayesian techniques like MCMC and {\sc MultiNest} have been designed specifically to explore the Bayesian posterior
distributions, they have also been used to obtain (approximate) profile likelihoods~\cite{2007JHEP...08..023A,2008JHEP...12..024T,Roszkowski:2009sm}.

{\sc MultiNest} \cite{feroz08, multinest}, a publicly available implementation of the nested sampling algorithm, has been shown to reduce the
computational cost of performing Bayesian analysis typically by two orders of magnitude as compared with basic MCMC techniques. {\sc MultiNest} has
been integrated in the \texttt{SuperBayeS} code\footnote{Available from: \texttt{www.superbayes.org}} for fast and efficient exploration of SUSY
models. Recently, it has been demonstrated (at least for a limited region of the CMSSM parameter space around a specific benchmark point) that neural networks techniques can be used to reduce the
computational efforts for these parameter scans by an additional factor of $\sim 10^4$~\cite{2011JHEP...03..012B}. 


For highly non-Gaussian problems like supersymmetric parameter determination, inference can depend strongly on whether one
chooses to work with the posterior distribution (Bayesian) or profile likelihood
(frequentist)~\cite{2007JHEP...08..023A,2008JHEP...12..024T,Roszkowski:2009sm}. There is a growing consensus that both the posterior
and the profile likelihood ought to be explored in order to obtain a fuller picture of the statistical constraints from
present-day and future data. This begs the question, which we address in this paper, of the algorithmic solutions available to
reliably explore both the posterior and the profile likelihood in the context of SUSY phenomenology.

Recently, a genetic algorithm (GA) based method was
developed in \cite{2010JHEP...04..057A} specifically to obtain the
profile likelihoods for CMSSM parameters and the resultant
distributions were then compared with the profile likelihoods obtained
with {\sc MultiNest}, run in the standard configuration commonly used in the literature. The authors found that {\sc MultiNest} missed
several high likelihood regions in the CMSSM parameter space making
the resultant profile likelihood functions highly inaccurate.  They
further questioned the accuracy of posterior distributions obtained
with {\sc MultiNest}. They went on to argue that {\sc MultiNest} might not be able to
find these high likelihood regions even if its termination criterion
is adjusted for finding the profile likelihoods. One of the aims of  this paper
is to show that these shortcomings can be avoided through proper
configuration of the {\sc MultiNest}. 

The outline of this paper is as follows. In Sec. \ref{sec:stats} we first give an introduction to the Bayesian and frequentist frameworks for statistical
analysis. We briefly describe the {\sc MultiNest} algorithm and how it can be tuned for evaluating profile likelihoods in Sec.~\ref{sec:multinest}. In
Sec.~\ref{sec:toymodel} we use two relevant toy problems to study the ability of {\sc MultiNest} to reconstruct  both the posterior and profile likelihood.
We then obtain the profile likelihood function of CMSSM parameters using {\sc MultiNest} in Sec.~\ref{sec:cmssm} and compare our results with the GA method
adopted in \cite{2010JHEP...04..057A}. We also discuss implications for direct and indirect detection prospects. Finally we present our conclusions in Sec.~\ref{sec:conclusions}.

%
\section{Statistical Framework}\label{sec:stats}
%
%
%
\subsection{Bayesian Modelling}\label{sec:stats:bayesian}

We briefly recall some basics about Bayesian inference, referring the reader to e.g.~\cite{Trotta:2008qt} for further details. Bayesian inference is based on Bayes' theorem,  stating that
\begin{equation} 
\Pr(\mathbf{\Theta}|\mathbf{D}) =
\frac{\Pr(\mathbf{D}|\,\mathbf{\Theta})\Pr(\mathbf{\Theta})}{\Pr(\mathbf{D})},
\label{eq:bayes}
\end{equation}
where $\Pr(\mathbf{\Theta}|\mathbf{D})$ is the posterior probability distribution of
the parameters $\mathbf{\Theta}$ given data $\mathbf{D}$, $\Pr(\mathbf{D}|\mathbf{\Theta}) \equiv \mathcal{L}(\mathbf{\Theta})$ is the likelihood function,
$\Pr(\mathbf{\Theta})$ is the prior, and $\Pr(\mathbf{D}) \equiv \mathcal{Z}$ is
the Bayesian evidence (or model likelihood). The latter is a normalization constant, obtained by averaging the likelihood over the prior:
\begin{equation}
\mathcal{Z} =
\int{\mathcal{L}(\mathbf{\Theta})\Pr(\mathbf{\Theta})}d\mathbf{\Theta}.
\label{eq:evidence}
\end{equation}
In Bayesian statistics, the $N$--dimensional posterior distribution described in Eq.~\eqref{eq:bayes} constitutes
the complete Bayesian inference of the parameter values. The inference about an individual parameter $\theta_{\rm
i}$ is then given by the marginalised posterior distribution $P(\theta_{\rm i}| {\bf D})$ which can be obtained by
integrating (marginalizing) the $N$--dimensional posterior distribution over all the parameters apart from
$\theta_{\rm i}$.
If samples from the full posterior are available (having been generated for example via Markov Chain Monte Carlo sampling), then the above integral is replaced by a simple counting procedure: since posterior samples are distributed according to the posterior, the marginal posterior for $\theta_{\rm i}$ above is obtained from the full posterior by dividing the range of $\theta_{\rm i}$ in a number of bins and counting how many posterior samples fall in each bin. 
A 1-D marginal posterior interval corresponding to a symmetric credible region containing $1-\alpha$ of posterior probability for $\theta_{\rm i}$ is delimited by an interval 
 $[\theta_{\rm i}^-, \theta_{\rm i}^+]$ such that:
\begin{eqnarray}
 \int_{-\infty}^{\theta_{\rm i}^-} P(\theta_{\rm i}|{\bf D}) d\theta_{\rm i} = \alpha/2 \quad \text{ and } 
  \int_{\theta_{\rm i}^+}^{\infty} P(\theta_{\rm i}|{\bf D}) d\theta_{\rm i} = \alpha/2.
\label{eq:bayes_credible_region}
\end{eqnarray}
\subsection{Profile Likelihood-Based Frequentist Analysis}\label{sec:stats:frequentist}

In the classical or frequentist school of statistics, one defines the
probability of an event as the limit of its frequency in a large
number of trials.  Classical confidence intervals based on the Neyman
construction are defined as the set of parameter points in which some
real-valued function, or \textit{test statistic}, $t$ evaluated on the
data falls in an acceptance region $W_\mathbf{\Theta} = [t_-, t_+]$.
If the boundaries of the acceptance regions are such that $P( t\in
W_\mathbf{\Theta}\, | \, \mathbf{\Theta}) \ge 1-\alpha$ is satisfied,
then the resulting intervals will cover the true value of $\mathbf \Theta$
with a probability of at least $1-\alpha$.  Likelihood ratios are
often chosen as the test statistic on which frequentist intervals are
based.  When $\mathbf{\Theta}$ is composed of parameters of interest,
$\theta$, and nuisance parameters, $\mbox{$\psi$}$, a common choice of test
statistic is the profile likelihood ratio
\begin{equation}
\lambda(\theta) \equiv \frac{\mathcal{L}(\theta, \hat{\hat{\mbox{$\psi$}}})}{\mathcal{L}(\hat{\theta}, \hat{\mbox{$\psi$}})}.
\label{eq:profile_like}
\end{equation}
where $\hat{\hat{\mbox{$\psi$}}}$ is the conditional maximum likelihood estimate
(MLE) of $\mbox{$\psi$}$ with $\theta$ fixed and $\hat{\theta}, \hat{\mbox{$\psi$}}$ are
the unconditional MLEs.  Under certain regularity conditions, Wilks
showed that the distribution of $-2\ln\lambda(\theta)$ converges to a
chi-square distribution with a number of degrees of freedom given by
the dimensionality of $\theta$~\cite{Wilks}.  Thus, by assuming that the asymptotic
distribution is a good approximation of the finite sample case, one
can trivially define the acceptance region $W_\mathbf{\Theta}$ from
standard lookup tables.  One must keep in mind that these intervals
based on the asymptotic properties of the profile likelihood ratio
require certain regularity conditions to be fulfilled and are not
guaranteed to have strict coverage.  In particular, in cases with
complex multimodal likelihoods one might reasonably suspect that the
distribution of $-2\ln\lambda(\theta)$ is still far from converging to
its asymptotic form (see \cite{2011JHEP...03..012B} for an illustration).
 While there exist modifications to the profile likelihood ratio that converge more rapidly~\cite{Bartlett,CoxReid,BarndorffNielsen}, within particle physics the profile likelihood ratio is a standard technique\footnote{In the following, we will refer to the profile likelihood ratio as ``profile likelihood'' or ``likelihood function'' for brevity.}.

%
%

One can approximate the $\lambda(\theta)$ using an arbitrary sampling of the likelihood function as long as one has access to the value of the likelihood function at the sampled points.  The procedure is simple: one groups the samples in bins of $\theta$ and for each bin searches for the maximum likelihood sample in that bin, which corresponds to $\mathcal{L}(\theta, \hat{\hat{\mbox{$\psi$}}})$.  

It can be seen from Eq.~\eqref{eq:profile_like} that the profile likelihood function requires the conditional MLEs $\hat{\hat{\mbox{$\psi$}}}(\theta)$ to be
found for each value of $\theta$ as well as the unconditional MLE $(\hat{\theta}, \hat{\mbox{$\psi$}})$, which represents the global maximum likelihood
point.  If the unconditional MLE is not properly resolved, this will effectively result in a different threshold on $-2\ln\lambda(\theta)$ corresponding to a
wrong confidence level.  Even if the unconditional MLE is found, we might also worry how well the conditional MLEs $\hat{\hat{\mbox{$\psi$}}}(\theta)$ are
found, particularly when $\mathcal{L}(\theta, \hat{\hat{\mbox{$\psi$}}}) \ll \mathcal{L}(\hat{\theta}, \hat{\mbox{$\psi$}})$.


\section{Using {\sc MultiNest} for Profile Likelihood Exploration}\label{sec:multinest}


Nested sampling \cite{skilling04} is a Monte Carlo method whose primary aim is the efficient calculation of the Bayesian evidence given by Eq.~\eqref{eq:evidence}  As a by-product, the algorithm also produces posterior samples which can be used to map out the posterior distribution of Eq.~\eqref{eq:bayes}. Those same samples have also been used to estimate the profile likelihood, a point to which we return below in greater detail. Nested sampling calculates the evidence by transforming the
multi--dimensional evidence integral into a one--dimensional integral that is easy to evaluate numerically. This
is accomplished by defining the prior volume $X$ as $dX = \Pr(\mathbf{\Theta})d \mathbf{\Theta}$, so that
\begin{equation}
X(\Lambda) = \int_{\mathcal{L}\left(\mathbf{\Theta}\right) > \Lambda} \Pr(\mathbf{\Theta}) d\mathbf{\Theta},
\label{eq:Xdef}
\end{equation}
where the integral extends over the region(s) of parameter space contained within the iso-likelihood contour
$\mathcal{L}(\mathbf{\Theta}) = \Lambda$. The evidence integral, Eq.~(\ref{eq:evidence}), can then be written as:
\begin{equation}
\mathcal{Z}=\int_{0}^{1}{\mathcal{L}(X)}dX,
\label{eq:nest_evidence}
\end{equation}
where $\mathcal{L}(X)$, the inverse of Eq.~(\ref{eq:Xdef}), is a  monotonically decreasing function of $X$. 
Thus, if one can evaluate the likelihoods $\mathcal{L}_{i}=\mathcal{L}(X_{i})$, where $X_{i}$ is a sequence of
decreasing values,
\begin{equation}
0<X_{M}<\cdots <X_{2}<X_{1}< X_{0}=1,
\end{equation}
the evidence can be approximated numerically using standard quadrature methods as a weighted sum
\begin{equation}
\mathcal{Z}={\textstyle {\displaystyle \sum_{i=1}^{M}}\mathcal{L}_{i}w_{i}},
\label{eq:NS_sum}
\end{equation}
where the weights $w_{i}$ for the simple trapezium rule are given by $w_{i}=\frac{1}{2}(X_{i-1}-X_{i+1})$.

The summation in Eq.~(\ref{eq:NS_sum}) is performed as follows. The iteration counter is first set to~$i=0$ and
$n_{\rm live}$ `active' (or `live') samples are drawn from the full prior $\Pr(\mathbf{\Theta})$, so the initial
prior volume is $X_{0} = 1$. The samples are then sorted in order of their likelihood and the smallest (with
likelihood $\mathcal{L}_{0}$) is removed from the active set (hence becoming `inactive') and replaced by a point
drawn from the prior subject to the constraint that the point has a likelihood $\mathcal{L}>\mathcal{L}_{0}$. The
corresponding prior volume contained within the iso-likelihood contour associated with the new live point will be
a random variable given by $X_{1} = t_{1} X_{0}$, where $t_{1}$ follows the distribution $\Pr(t) = Nt^{N-1}$
(i.e., the probability distribution for the largest of $N$ samples drawn uniformly from the interval $[0,1]$). At
each subsequent iteration $i$, the removal of the lowest likelihood point $\mathcal{L}_{i}$ in the active set,
the drawing of a replacement with $\mathcal{L} > \mathcal{L}_{i}$ and the reduction of the corresponding prior
volume $X_{i}=t_{i} X_{i-1}$ are repeated, until the entire prior volume has been traversed. The algorithm thus
travels through nested shells of likelihood as the prior volume is reduced.

The nested sampling algorithm is terminated when the evidence has been computed to a pre-specified precision. The
evidence that could be contributed by the remaining live points is estimated as $\Delta{\mathcal{Z}}_{\rm i} =
\mathcal{L}_{\rm max}X_{\rm i}$, where ${\cal L}_{\rm max}$ is the maximum-likelihood value of the remaining live
points, and $X_i$ is the remaining prior volume. The algorithm terminates when $\Delta{\mathcal{Z}}_{\rm i}$ is
less than a user-defined value.

The most challenging task in implementing nested sampling is to draw samples from the prior within the hard
constraint $\mathcal{L}> \mathcal{L}_i$ at each iteration $i$. The {\sc MultiNest} algorithm~\cite{feroz08, multinest}  tackles this problem
through an ellipsoidal rejection sampling scheme. The  live point set is enclosed within a set of (possibly
overlapping)  ellipsoids and a new point is then drawn uniformly from the region enclosed by these ellipsoids.
The ellipsoidal decomposition of the live point set is chosen to minimize the sum of volumes of the ellipsoids.
The ellipsoidal decomposition is well suited to dealing with posteriors that have curving degeneracies, and
allows mode identification in multi-modal posteriors. If there are subsets of the ellipsoid set that do not
overlap with the remaining ellipsoids, these are identified as a distinct mode and subsequently evolved
independently.

The two most important parameters  that control the parameter space exploration in nested sampling are the number of live points $n_{\rm live}$ -- which determines the resolution at which the parameter space is explored --
and a tolerance parameter $\mathrm{tol}$, which defines the termination criterion based on the accuracy of the evidence. As discussed in the
previous section, evaluating profile likelihoods is much more challenging than evaluating posterior
distributions. Therefore, one should not expect that a vanilla setup for {\sc MultiNest} (which is adequate for an accurate exploration of the posterior distribution) will automatically be optimal for profile likelihoods evaluation. Generally, a larger number of live points is necessary to explore profile likelihoods accurately. Moreover, setting
$\mathrm{tol}$ to a smaller value results in {\sc MultiNest} gathering a larger number of samples in the high likelihood regions (as termination is delayed). This is usually not necessary for the posterior distributions, as the prior volume occupied by high
likelihood regions is usually very small and therefore these regions have relatively small probability mass. For
profile likelihoods, however, getting as close to the true global maximum is crucial and therefore one should set
$\mathrm{tol}$ to a relatively smaller value. We now investigate these issues in the context of two relevant toy problems.

\section{Application to Toy Models}\label{sec:toymodel}

In order to demonstrate the differences between profile likelihoods and the posterior distributions reconstruction with {\sc MultiNest}, we consider
two 8-D toy problems in this section. The dimensionality of these toy problems is chosen to correspond to the
dimensionality of CMSSM analyses in which 4 CMSSM parameters are varied along with 4 Standard Model (SM)
parameters (see Sec.~\ref{sec:cmssm}). 

\begin{figure}
\begin{center}
\subfigure[]{\includegraphics[width=0.48\columnwidth]{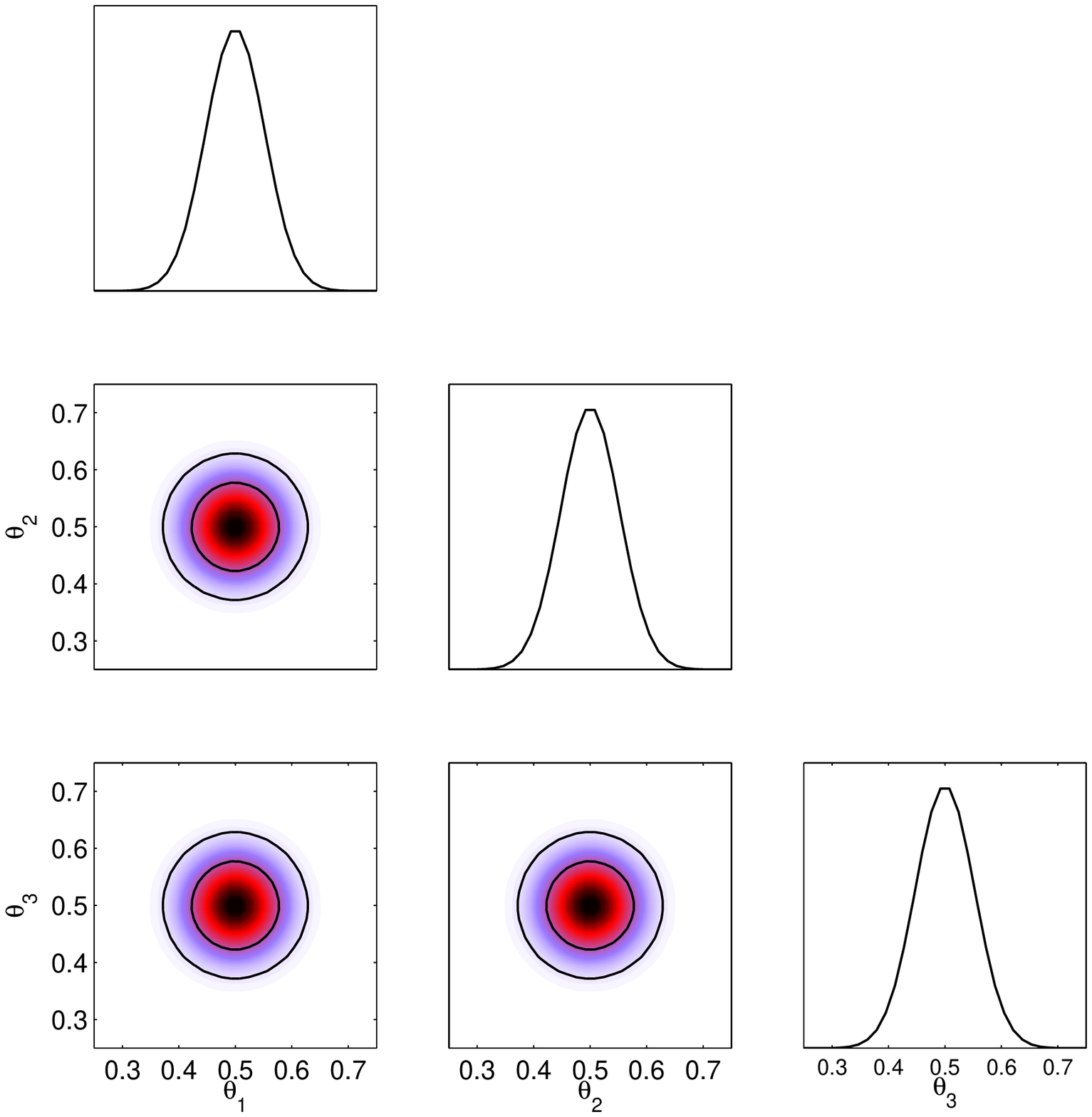}}\hspace{0.1cm}
\subfigure[]{\includegraphics[width=0.48\columnwidth]{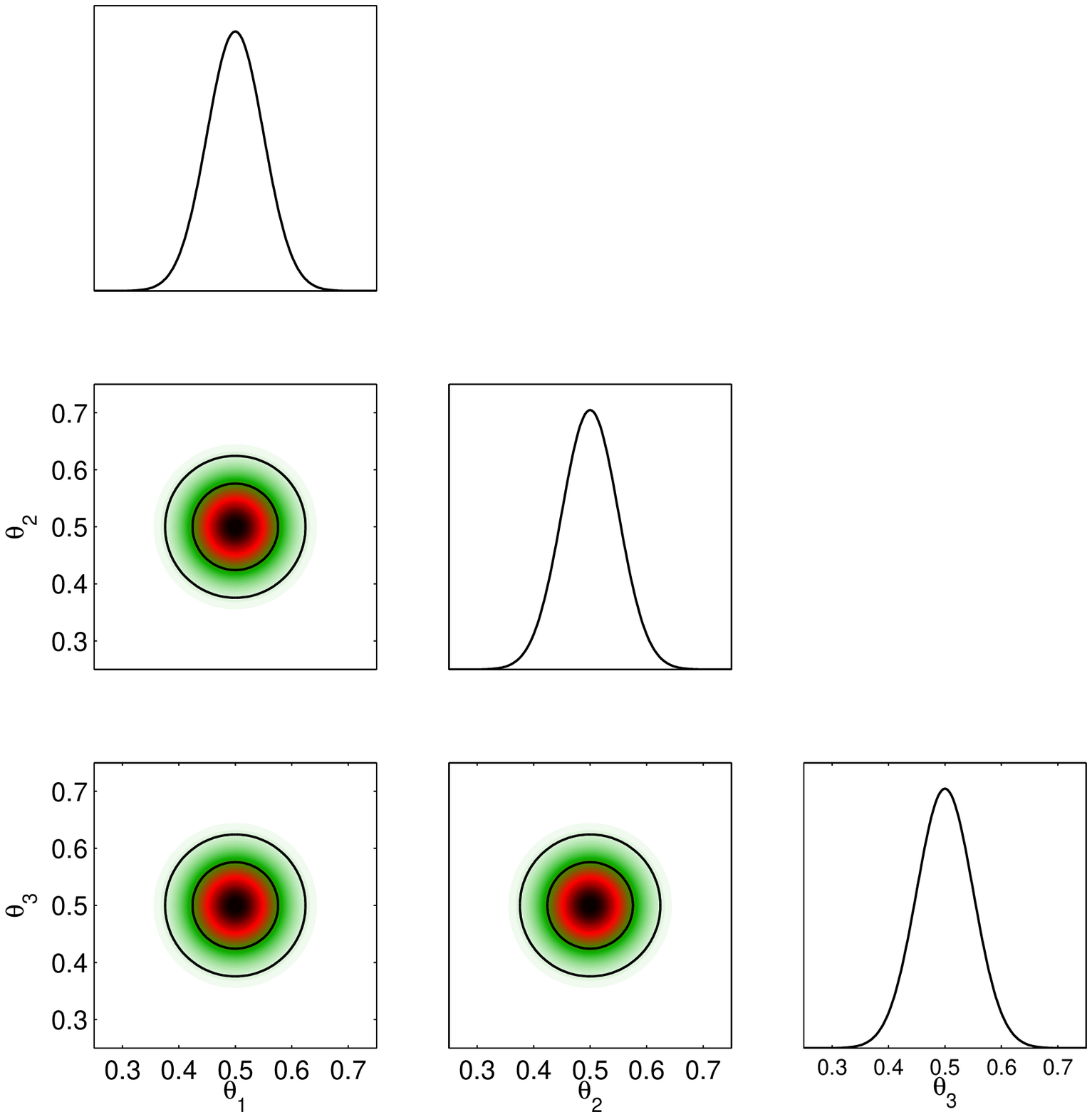}}
\caption{Analytical 1-D and 2-D (a) marginalized posterior distributions and (b)
  profile likelihood functions for the unimodal 8-D Gaussian
  problem. The contours represent the 68\% and 95\% Bayesian credible
  regions in (a) and profile likelihood confidence intervals in (b)
  respectively.} \label{fig:umodal_analytical}
\end{center}
\end{figure}

The first problem we consider is a uni-modal 8-D isotropic Gaussian with the following likelihood function:
\begin{equation}
\mathcal{L}(\Theta) = \frac{1}{(2\pi)^4 \sigma^8} \exp\left[ -\frac{1}{2} \sum_{\rm j = 1}^{8} \frac{\left(\theta_{\rm j} - \mu_{\rm j}\right)^2}{\sigma^2} \right]
\end{equation}
with $\sigma = 0.05$ and $\mu_{\rm j} = 0.5$ ($j=1, \dots,8)$. We assume uniform priors
$\mathcal{U}(0,1)$ for all 8 parameters.  The analytical posterior
distributions and profile likelihood functions are shown in
Fig.~\ref{fig:umodal_analytical} for a subset of the parameters. As expected, the posterior and
profile likelihood distributions are identical in this case.

\begin{figure}
\begin{center}
\includegraphics[width=1\columnwidth]{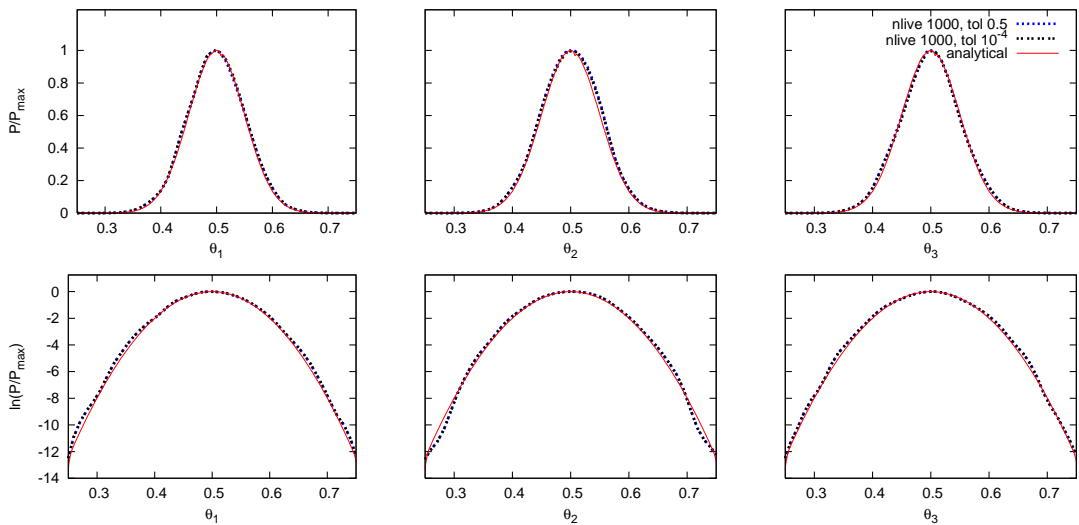}
\caption{1-D posterior distributions for the first 3 parameters of the
  uni-modal 8-D Gaussian problem. The red curve shows the analytical
posterior distribution while the blue dashed and grey dotted curves show the posterior distributions recovered by
{\sc MultiNest} with $\mathrm{tol}$ set to 0.5 and $1 \times 10^{-4}$ respectively. 1000 live points were used in
both cases. The upper and lower panels show the posterior and $\ln$ of posterior distributions respectively.}
\label{fig:umodal_marg}
\end{center}
\end{figure}

\begin{figure}
\begin{center}
\includegraphics[width=1\columnwidth]{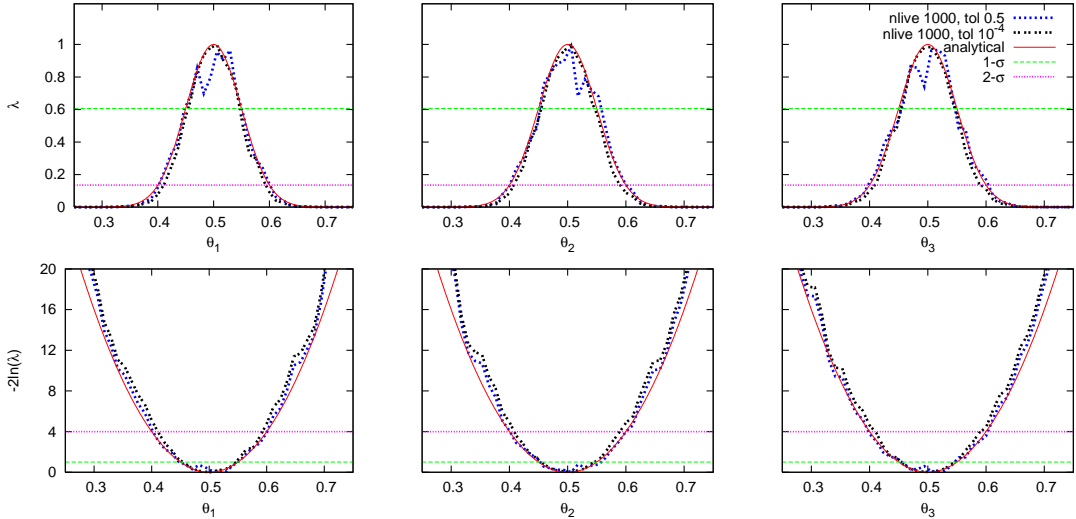}
\caption{1-D profile likelihood functions for the first 3 parameters
  of the uni-modal 8-D Gaussian problem. The red curve shows the
analytical profile likelihood functions while the blue dashed and grey dotted curves show the profile
likelihoods recovered by {\sc MultiNest} with $\mathrm{tol}$ set to 0.5 and $1 \times 10^{-4}$ respectively. 1000
live points were used in both cases. The upper and lower panels show the profile likelihood and $-2\ln$ of profile 
likelihood functions respectively. Green and magenta horizontal lines represent the $1\sigma$ and $2\sigma$ profile 
likelihood intervals respectively.}
\label{fig:umodal_profile}
\end{center}
\end{figure}

We now reconstruct both profile likelihood and posterior distribution with {\sc MultiNest}, using $n_{\rm live} = 1000$ with
$\mathrm{tol} = 0.5$ (which is expected to be adequate for posterior reconstruction) and $1 \times 10^{-4}$ (a much lower value
targeted at profile likelihood reconstruction). The posterior distributions and profile likelihoods recovered by {\sc MultiNest
}for the first three parameters are shown in Figs.~\ref{fig:umodal_marg} and \ref{fig:umodal_profile}  respectively. It is
evident from these plots that recovered posterior distributions for both $\mathrm{tol} = 0.5$ and $1 \times 10^{-4}$ are almost
identical to the analytical posterior more than $3\sigma$ into the tails. However, the recovered profile likelihood for
$\mathrm{tol} = 0.5$ is quite noisy and inaccurate, especially around the peak. This is because {\sc MultiNest} has not explored
the high likelihood region in great detail, as the algorithm is terminated after about 163,000 likelihood evaluations, which is
adequate to calculate the evidence to sufficient accuracy. We can circumvent this problem by setting $\mathrm{tol} = 1 \times
10^{-4}$ (dotted blue curves), which results in a larger number of likelihood evaluations, around 261,000. The situation around
the peak is now greatly improved, although the tails of the profile likelihood are still not explored to very high accuracy
beyond the $\sim 3\sigma$ region. Nevertheless, the overall profile likelihood has been recovered with sufficient accuracy to
allow for a reasonably correct parameter estimation.

\begin{figure}
\begin{center}
\subfigure[]{\includegraphics[width=0.48\columnwidth]{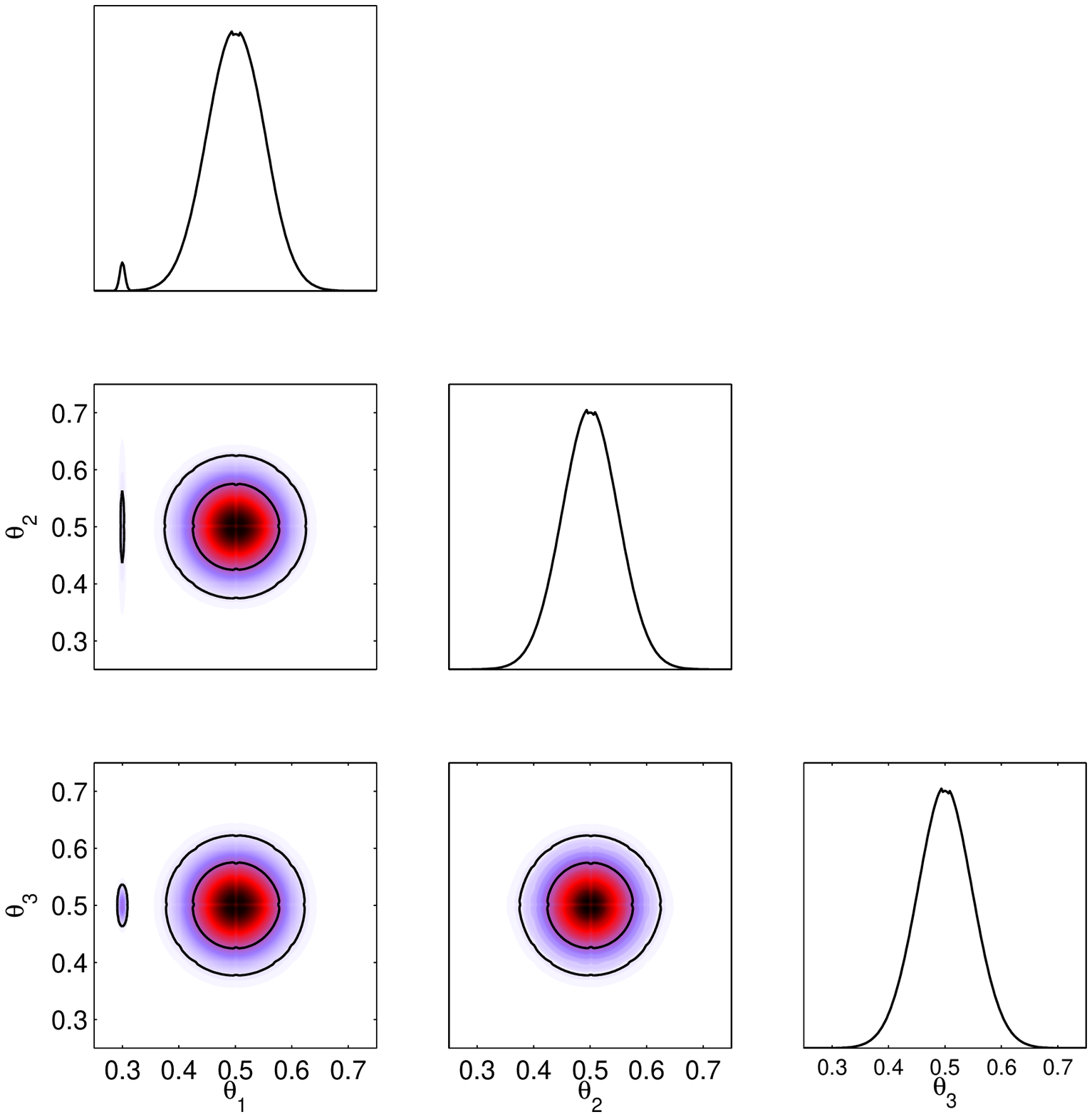}}\hspace{0.1cm}
\subfigure[]{\includegraphics[width=0.48\columnwidth]{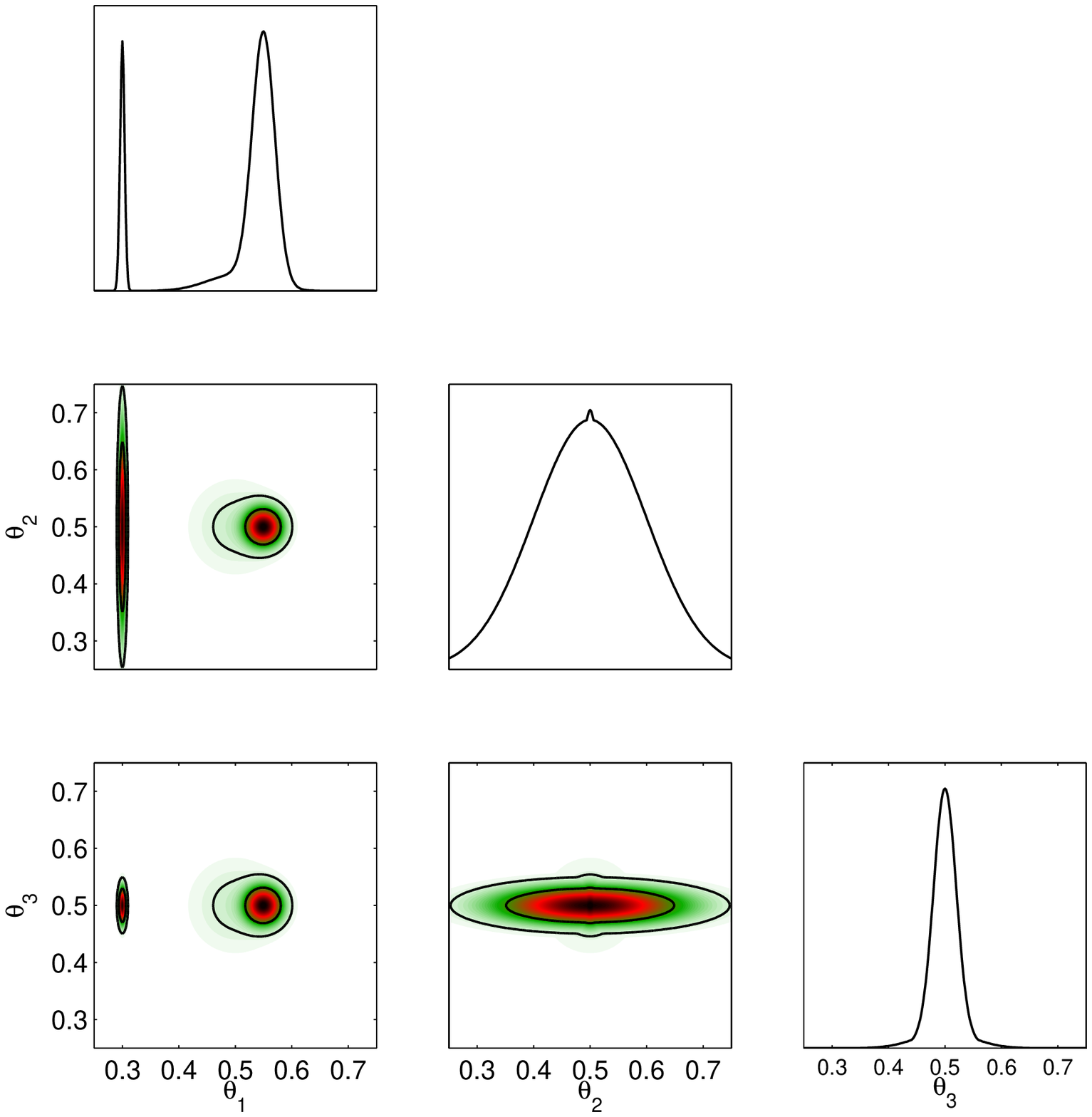}}
\caption{1-D and 2-D (a) marginalized posterior distributions and (b) profile likelihood functions for the 8-D Gaussian
mixture problem. The contours represent the 68\% and 95\% Bayesian
credible regions in (a) and 
profile likelihood confidence intervals in (b)
respectively. The plots were generated after
dividing the analytical posterior values in $200$ bins so that a fair
comparison can be made with the distributions obtained using {\sc
  MultiNest}. } \label{fig:mmodal_analytical}
\end{center}
\end{figure}

The second problem we consider is a multi-modal 8-D Gaussian mixture model with the
following likelihood function:
\begin{equation}
\mathcal{L}(\Theta) = \sum_{\rm i = 1}^{3} w_{\rm i} \frac{1}{(2\pi)^4 \prod_{\rm j = 1}^{8} \sigma_{\rm ij}}
\exp\left[ -\frac{1}{2} \sum_{\rm j = 1}^{8} \frac{\left(\theta_{\rm i} - \mu_{\rm ij}\right)^2}{\sigma_{\rm
ij}^2} \right]
\end{equation}
with the following values for the parameters defining the likelihood function:
\begin{align}
\mathcal{M}_1: & \, w_1 = 0.98, \sigma_{\rm 1i} = 0.05 \text{ and } \mu_{\rm 1i} = 0.5 \, (1 \le i \le 8); \\
\mathcal{M}_2: & \,w_2 = 0.01, \sigma_{\rm 2i} = 0.02\, (1 \le i \le 8), \mu_{21} = 0.55, \mu_{\rm 2i} =
0.5 \, (2 \le i \le 8); \\ 
\mathcal{M}_3: & \, w_3 = 0.01, \sigma_{31} = 0.004, \sigma_{32} = 0.1, \sigma_{\rm 3i} = 0.02 \, (3 \le i \le 8), \nonumber \\
& \mu_{31} = 0.3, \mu_{\rm 3i} = 0.5\, (2 \le i \le 8). 
\end{align}
We assume uniform priors $\mathcal{U}(0,1)$ for all 8 parameters. This describes a multi-modal distribution with the broad
Gaussian of the previous example centered in the middle of the parameter space. A narrow isotropic Gaussian lies slightly to the
right of the broad Gaussian in $\theta_1$ while another narrow Gaussian lies to the left. We denote these three modes by
$\mathcal{M}_1$, $\mathcal{M}_2$ and $\mathcal{M}_3$ respectively.  $\mathcal{M}_3$ has its variances in the first two
dimensions chosen in such a way to mimic the funnel region at low $m_{1/2}$ values in the CMSSM model. The weights of the mixture model are set such
that the $\mathcal{M}_1$ occupies 98\% of the posterior probability mass while $\mathcal{M}_2$ and $\mathcal{M}_3$ collectively
occupy only 2\% of the probability mass and therefore we would expect the posterior probability distribution to be dominated by
$\mathcal{M}_1$. The analytical 1-D and 2-D posterior distributions and profile likelihood functions are shown in
Fig.~\ref{fig:mmodal_analytical} (only distributions of the first 3 parameters are shown, as parameters $\theta_4$ onwards
behave exactly as $\theta_3$). As expected, the posterior distributions are dominated almost entirely by $\mathcal{M}_1$, the
mode with the largest posterior probability mass.  $\mathcal{M}_3$ barely registers in the posterior distribution as the
secondary peak in the $\theta_1$ direction, while $\mathcal{M}_2$ is completely masked by $\mathcal{M}_1$.

The profile likelihood, on the other hand, is dominated by the narrow, highly-peaked modes $\mathcal{M}_2$ and $\mathcal{M}_3$,
whose peak likelihood values are 9.25 and 9.61 times higher, respectively, than the likelihood value at the peak of
$\mathcal{M}_1$. This is indeed reflected in the right panels of Fig.~\ref{fig:mmodal_analytical}. The 1-D profile likelihood
for $\theta_1$ is bi-modal with $\mathcal{M}_2$ and $\mathcal{M}_3$ constituting the two modes. $\mathcal{M}_3$ only produces a
heavier, non-Gaussian tail extending downwards from $\theta_1 = 0.55$ (where the maximum likelihood peak of $\mathcal{M}_2$ is
located).  Looking instead at the profile likelihood for $\theta_2$, the (very subdominant) contribution from $\mathcal{M}_1$
barely shows up as a small extra peak on top of the broader peak from $\mathcal{M}_3$  at $\theta_2 = 0.5$. The profile
likelihood for $\theta_2$ consists almost entirely of $\mathcal{M}_3$ as its standard deviation in $\theta_2$ is 5 times larger
than the one of $\mathcal{M}_2$ while the likelihood value at the peak of $\mathcal{M}_3$ is only slightly smaller than the
value for $\mathcal{M}_2$. We notice that confidence regions for the profile likelihood plotted in the 2-D panels are only
approximately correct, as we have assumed that the likelihood distribution follows a multivariate Gaussian distribution in order
to derive the confidence limits as discussed in Sec.~\ref{sec:stats:frequentist}, which is obviously not correct for this
particular toy model.

\begin{figure}
\begin{center}
\includegraphics[width=1\columnwidth]{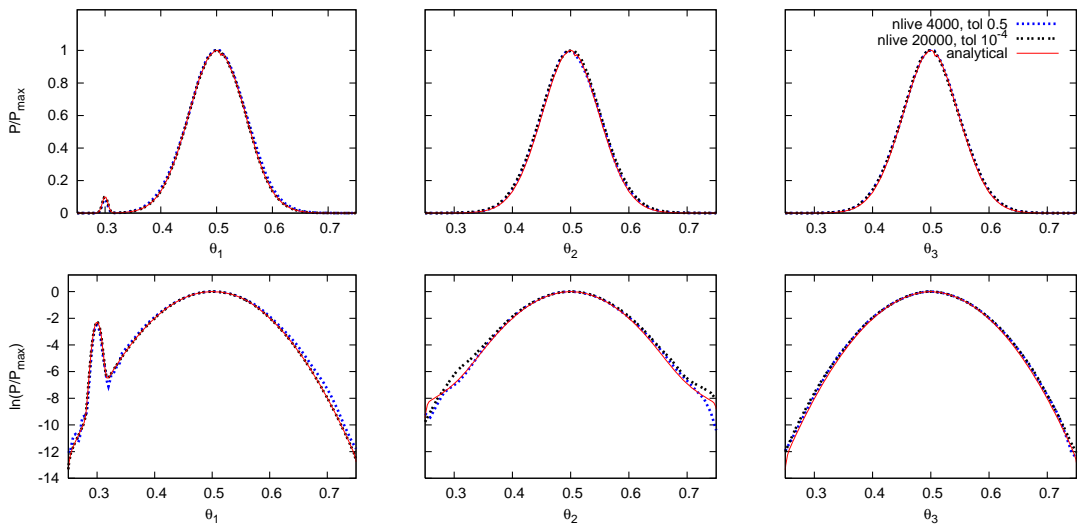}
\caption{1-D posterior distributions for the first 3 parameters of the
  8-D Gaussian mixture problem. The red curve shows the analytical
posterior distribution while the blue dashed and grey dotted curves show the posterior distributions recovered by
{\sc MultiNest} with $n_{\rm live}=4,000$, $\mathrm{tol} = 0.5$ and $n_{\rm live}=20,000$, $\mathrm{tol} = 1 \times
10^{-4}$ respectively. The upper and lower panels show the posterior and $\ln$ of posterior distributions 
respectively.}
\label{fig:mmodal_marg}
\end{center}
\end{figure}

\begin{figure}
\begin{center}
\includegraphics[width=1\columnwidth]{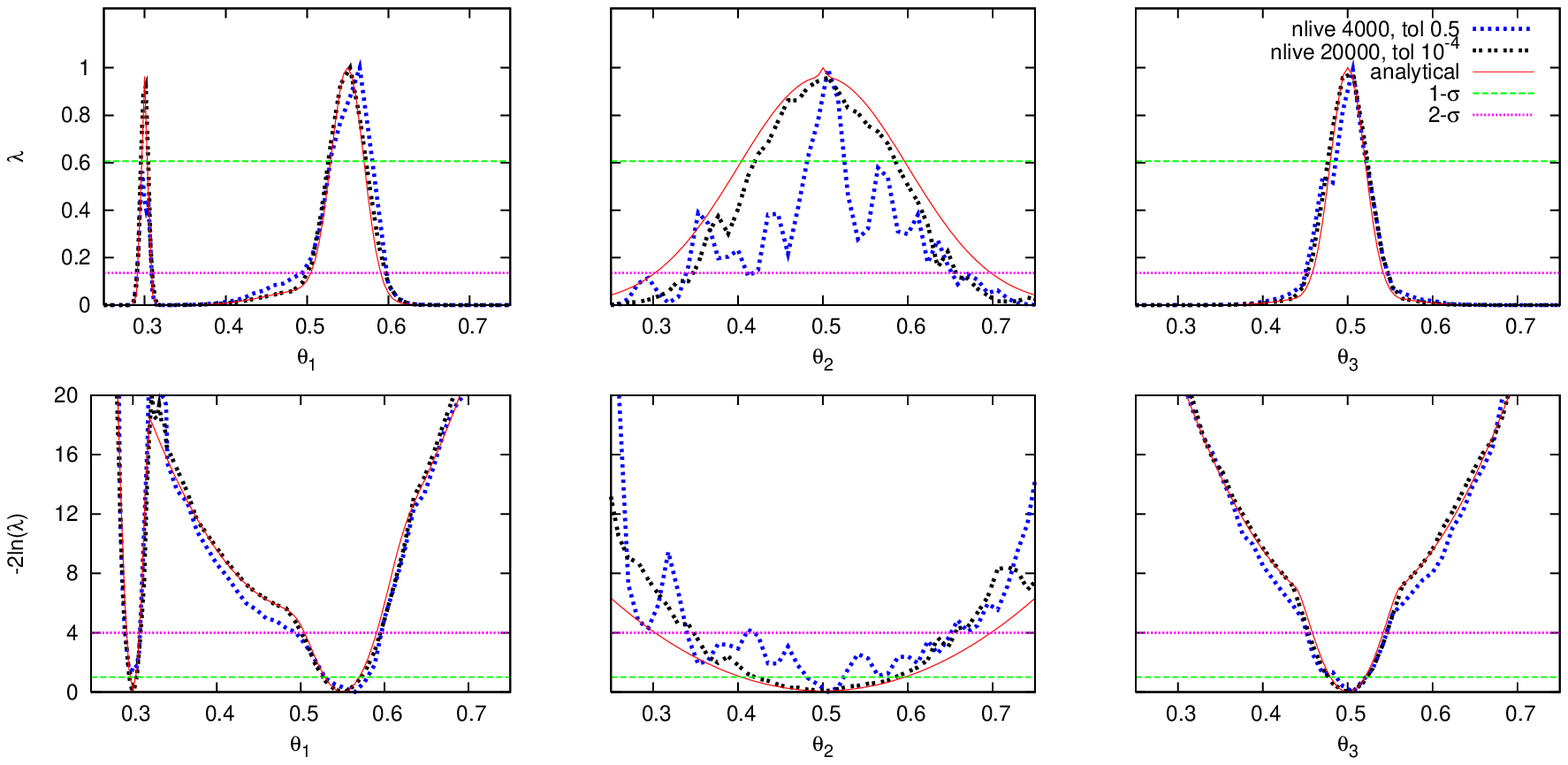}
\caption{1-D profile likelihood functions for the first 3 parameters
  of the 8-D Gaussian mixture problem. The red curve shows the
analytical profile likelihood function while the blue dashed and grey dotted curves show the profile
likelihoods recovered by {\sc MultiNest} with $n_{\rm live}=4,000$, $\mathrm{tol} = 0.5$ and $n_{\rm live}=20,000$,
$\mathrm{tol} = 1 \times 10^{-4}$ respectively. The upper and lower panels show the profile likelihood and $-2\ln$ 
of profile likelihood functions respectively. Green and magenta horizontal lines represent the $1\sigma$ and $2\sigma$ profile 
likelihood intervals respectively.}
\label{fig:mmodal_profile}
\end{center}
\end{figure}

These dramatic differences in the posterior and profile likelihood functions make this problem an extremely challenging one for
any numerical inference algorithm to tackle. Even just finding $\mathcal{M}_2$ and $\mathcal{M}_3$ is an extremely difficult
task as they are essentially spikes in the probability distribution. We show the recovered 1-D posterior distributions and
profile likelihood with {\sc MultiNest} in Figs.~\ref{fig:mmodal_profile} and \ref{fig:mmodal_marg},  respectively. With $n_{\rm
live} = 4,000$ and $\mathrm{tol} = 0.5$ (resulting in about 725,000 likelihood evaluations), {\sc MultiNest} finds all three Gaussians
and the recovered posterior distributions are almost identical to the analytical posterior distributions. However, the profile
likelihood functions are highly inaccurate and very noisy. This is because the high likelihood regions have not been explored in
great detail as the sampling proceeds according to the posterior mass, which is heavily concentrated in ${\mathcal
M}_1$. Therefore, the modes  ${\mathcal M}_2$ and ${\mathcal M}_3$ are explored with a proportionally lower number of live
points, as their posterior mass is small. 

This problem can be solved by running {\sc MultiNest} with $n_{\rm live} = 20,000$ and decreasing $\mathrm{tol}$ to $1 \times
10^{-4}$ which results in around 11 million likelihood evaluations. Increasing the number of live points ensures that
$\mathcal{M}_3$ encloses more live points and is explored at higher resolution. As can be seen from
Fig.~\ref{fig:mmodal_profile}, the recovered profile likelihood functions are much more accurate and less noisy with this setup.
The 1-D profile likelihoods for $\theta_1$ and $\theta_3$ are almost identical to the analytical ones, but we can see some
under-sampling of the profile likelihood for $\theta_2$. Despite this, just finding $\mathcal{M}_3$ is already an extremely
challenging task for any numerical technique, including GA, and although the recovered profile likelihood functions by {\sc
MultiNest} are not perfect, they are reasonably accurate, certainly within the $2\sigma$ confidence region.


\section{Application to the CMSSM}\label{sec:cmssm}

The Minimal Supersymmetric Standard Model (MSSM)~\cite{Farrar:1978xj, Dimopoulos:1981zb} with R-parity can solve the hierarchy
problem and provide a candidate dark matter (DM) particle. The MSSM with one particular choice of universal boundary conditions
at the grand unification scale, is called the Constrained Minimal Supersymmetric Standard Model
(CMSSM)~\cite{AlvarezGaume:1983gj}. In the CMSSM, the scalar mass $m_0$, gaugino mass $m_{1/2}$ and tri--linear coupling $A_0$
are assumed to be universal at a gauge unification scale $M_{\rm GUT} \sim 2 \times 10^{16}$ GeV. In addition, at the
electroweak scale one selects $\tan\beta$, the ratio of Higgs vacuum expectation values and sign$(\mu)$, where $\mu$ is the
Higgs/higgsino mass parameter whose square is computed from the potential minimisation conditions of electroweak symmetry
breaking (EWSB) and the empirical value of the mass of the $Z^0$ boson, $M_Z$. The family universality assumption is well
motivated since flavour changing neutral currents are observed to be rare. Indeed several string models (see, for example
Ref.~\cite{2006JHEP...06..029C, 1994NuPhB.422..125B}) predict approximate MSSM universality in the soft terms. 

The CMSSM has proved to be a popular choice for SUSY phenomenology because of the small number of free parameters and it has
recently been studied quite extensively in multi--parameter scans, both from a
frequentist~\cite{Ellis:2003si,Profumo:2004at,2009EPJC...64..391B} and a Bayesian
perspective~\cite{2004JHEP...10..052B,Allanach:2005kz,2006JHEP...05..002R,2008JHEP...12..024T, 2008JHEP...10..064F,
2007JHEP...08..023A,2010JHEP...04..057A,  2007JHEP...07..075R}. Bayesian and frequentist analyses are expected to give
consistent inferences for the CMSSM parameters if the data are sufficiently constraining. However, several groups have concluded
that we do not yet have enough constraining power in the available data to overridethe influence of priors in the Bayesian
framework. Furthermore, the structure of the parameter space is such that the likelihood presents ``spikes'' that contain little
posterior mass (for reasonable choice of priors). As a consequence, regions of high likelihood (favoured in a frequentist
analysis) do not necessarily match regions of large posterior mass (favoured under a Bayesian analysis), see e.g.
\cite{2008JHEP...12..024T}. Therefore, Bayesian and frequentist inferences should not be expected to yield quantitatively
consistent results. Both hurdles are expected to be overcome once ATLAS results become available, see ~\cite{Roszkowski:2009ye},
although the actual performance of different statistical approaches, as measured by their coverage properties, has only just
begun to be scrutinized~\cite{2011JHEP...03..012B, 2010arXiv1011.4297A}. 

Here we focus on the algorithmic problem of obtaining reliable posterior probability and profile likelihood maps with {\sc
MultiNest}. The authors of Ref.~\cite{2010JHEP...04..057A} claimed that the GA method produces more accurate profile likelihood
functions than {\sc MultiNest}. They also raised questions about the accuracy of the posterior distributions obtained with {\sc
MultiNest}. The previous section demonstrates that the posterior distributions from  {\sc MultiNest} are highly reliable, as
they match very well the analytical solution for both toy problems out to more than $3\sigma$ in the tails. We stress that this
is the case even for the less computationally intensive  {\sc MultiNest} configuration, with $n_{\rm live} = 1000$ and
$\mathrm{tol} = 0.5$.

We now reconstruct both the posterior distribution and the profile likelihood for the CMSSM parameters using {\sc MultiNest}, as
implemented in the \texttt{SuperBayeS v1.5} code in almost exactly the same
setup employed in Ref.~\cite{2010JHEP...04..057A}. The set of CMSSM and SM parameters together constitute an 8-dimensional
parameter space with $\Theta$ =($m_0$, $m_{1/2}$, $A_0$, $\tan\beta$, $m_t$, $m_b(m_b)^{\overline{MS}}$,
$\alpha^{\overline{MS}}$, $\alpha_s^{\overline{MS}}(M_Z))$ to be scanned and constrained with the presently available
experimental data. The ranges over which we explore these parameters are $m_0, m_{1/2} \in (0.05,~4)$ TeV, $A_0 \in (-4,~4)$
TeV, $\tan\beta \in (2,~65)$, $m_t \in(167.0,~178.2)$ GeV, $m_b(m_b)^{\overline{MS}} \in (3.92,~4.48)$ GeV,
$1/\alpha^{\overline{MS}}(M_Z) \in (127.835,~128.075)$ and $~\alpha_s^{\overline{MS}}(M_Z) \in (0.1096,~0.1256)$. We use a
slightly wider range for $\tan\beta$ and a smaller range for $A_0$ which was allowed to vary between $-7$ TeV and $7$ TeV in
\cite{2010JHEP...04..057A} as there is very little probability  mass as well as very few points with high likelihood with $|A_0|
> 4$ TeV.  The observables used in our analysis are exactly the same as given in Table~1 of Ref.~\cite{2010JHEP...04..057A}, in
order to allow a comparison.

\begin{figure}
\begin{center}
\subfigure[]{\includegraphics[width=0.8\columnwidth]{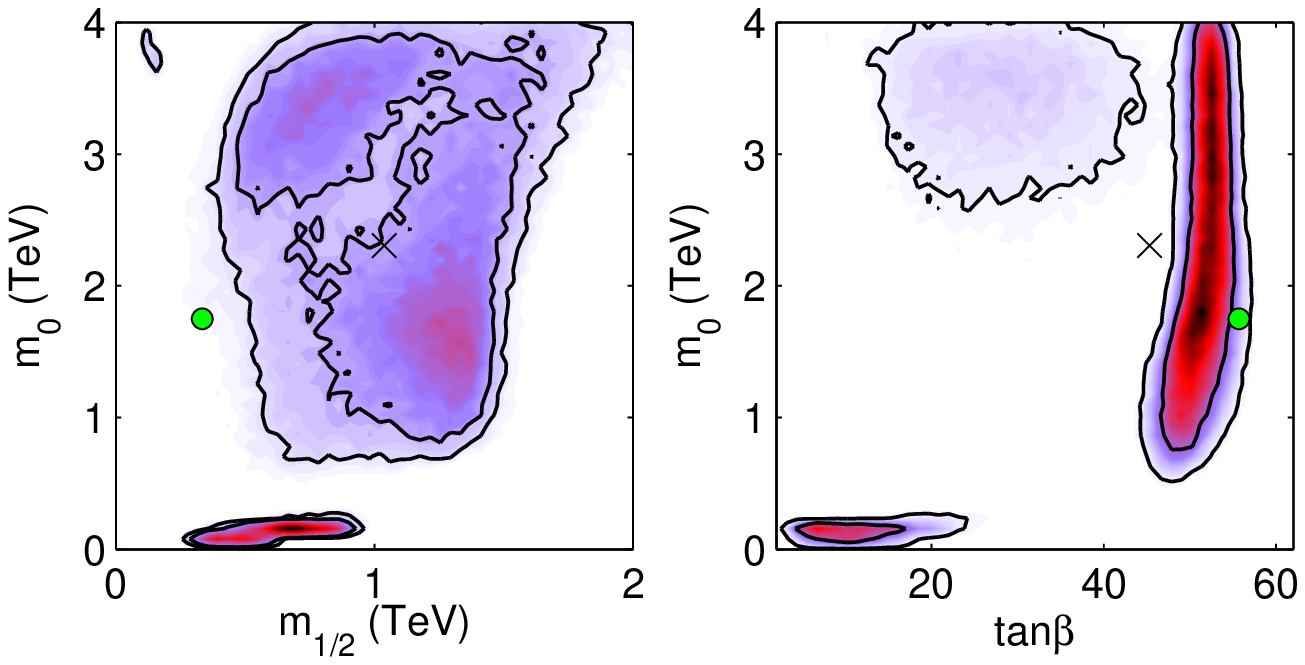}}\\
\subfigure[]{\includegraphics[width=0.8\columnwidth]{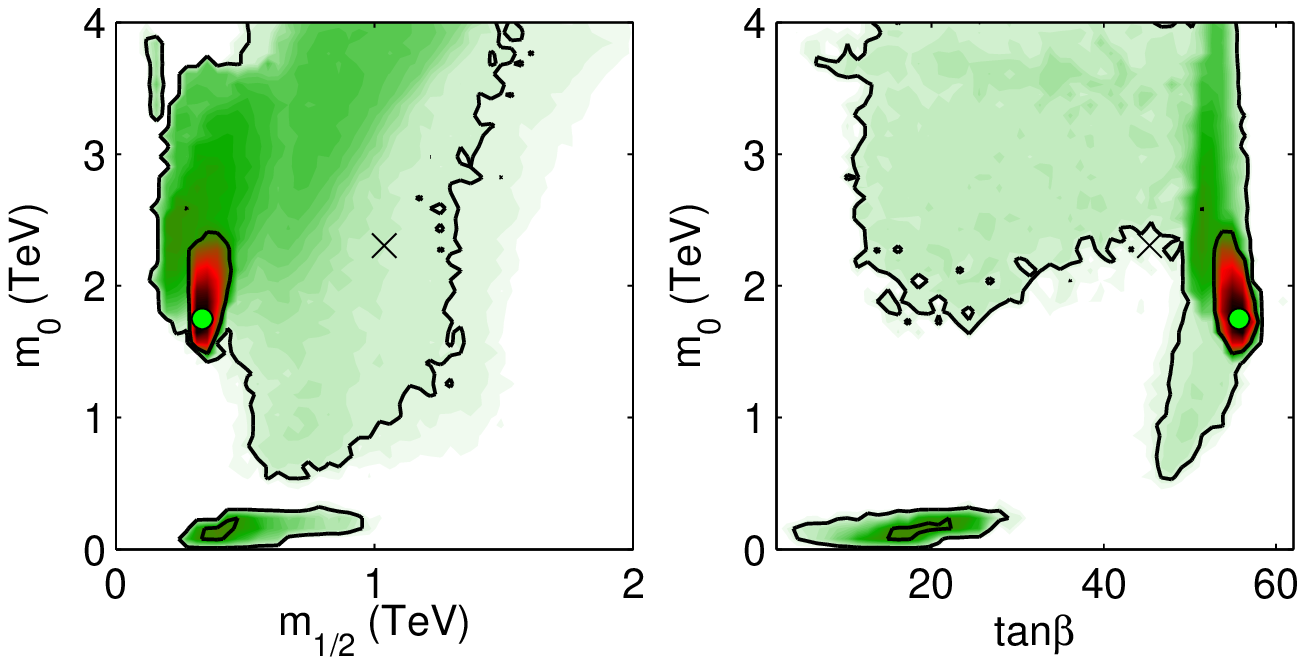}}
\caption{2-D (a) marginalized posterior distributions and (b) profile
  likelihoods for the CMSSM parameters using all presently available constraints. The contours represent
  the 68\% and 95\% Bayesian credible regions in (a) and profile
  likelihood confidence intervals in (b). Flat priors
  with 20,000 live points were used with $\mathrm{tol}$ set to $1
  \times 10^{-4}$ (no smoothing applied to the plots). The mean and best-fit parameters values are shown
  by a black cross and a green circle, respectively.}
\label{fig:cmssm_nlive_20k_tol_01}
\end{center}
\end{figure}

\begin{figure}
\begin{center}
\includegraphics[width=1\columnwidth]{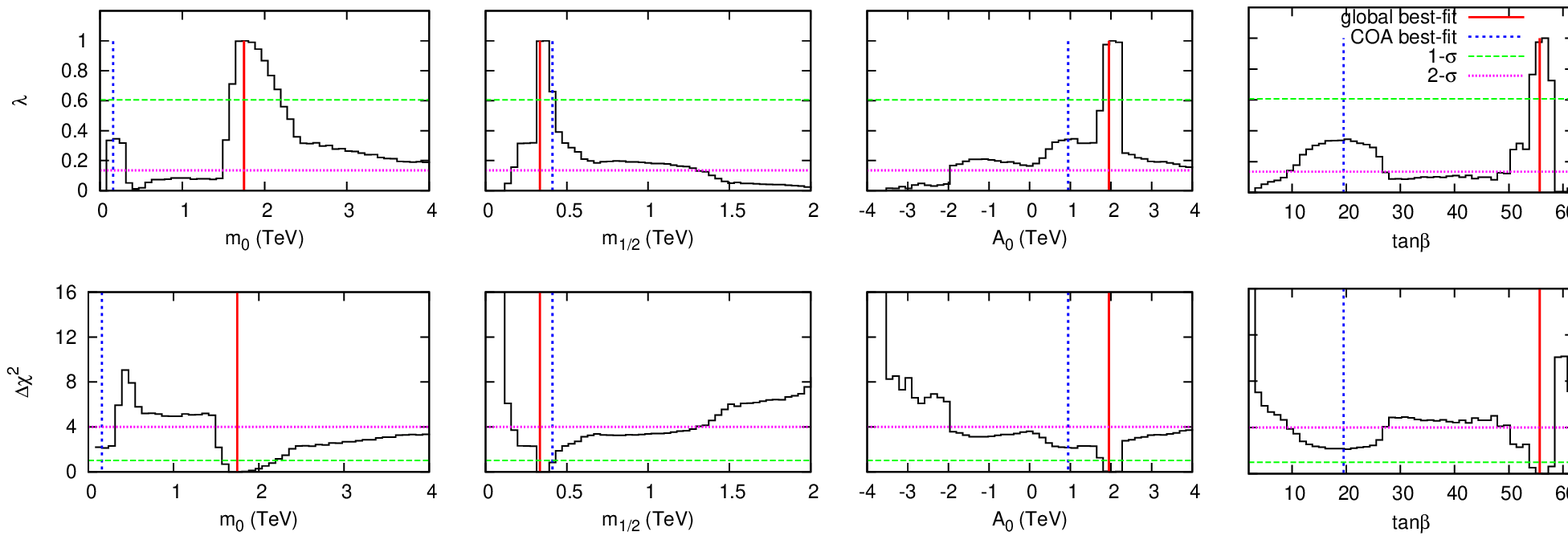}
\caption{1-D profile likelihoods for the CMSSM parameters normalized to the global best-fit point. The red solid and blue dotted
  vertical lines represent the global best-fit point ($\chi^2 = 9.26$, located in the focus point region) and the best-fit point
  found in the stau co-annihilation region ($\chi^2 = 11.38$) respectively. The upper and lower panel show the profile
  likelihood and $\Delta\chi^2$ values, respectively. Green (magenta) horizontal lines represent the $1\sigma$ ($2\sigma$)
  approximate confidence intervals. {\sc MultiNest} was run with flat priors, 20,000 live points and $\mathrm{tol}=1 \times
  10^{-4}$.}
\label{fig:cmssm_profile_1D}
\end{center}
\end{figure}

As discussed in Secs.~\ref{sec:multinest} and \ref{sec:toymodel}, a larger $n_{\rm live}$ and smaller $\mathrm{tol}$ should be used for profile
likelihood analyses with {\sc MultiNest}. We therefore used $n_{\rm live} = 20,000$ and $\mathrm{tol} = 1 \times 10^{-4}$. Flat priors were imposed on
all 8 parameters with parameter ranges described above. This resulted in about 5.5 million likelihood evaluations (compared to 3 million likelihood
evaluations for the GA method) taking 6 days on 10 2.4GHz CPUs and returned an evidence value of $\log(\mathcal{Z}) = -18.62 \pm 0.08$. The resultant
2-D marginalized posterior distributions and profile likelihoods in the $m_{1/2}-m_0$ and $\tan\beta-m_0$ planes are shown in
Fig.~\ref{fig:cmssm_nlive_20k_tol_01}. By comparing the posterior distribution in the upper left panel of Fig.~\ref{fig:cmssm_nlive_20k_tol_01} with
the upper left panel of Fig.~13 in~\cite{2008JHEP...12..024T}, obtained with the same setup but with  $n_{\rm live} = 4,000$ and $\mathrm{tol} = 0.5$,
it is apparent that the posterior is stable with respect to increasing the number of samples (Fig.~\ref{fig:cmssm_nlive_20k_tol_01} uses $\sim 20$
times more samples than the corresponding figure in Ref.~\cite{2008JHEP...12..024T}). Indeed, the evidence value obtained with this setup is
$\log(\mathcal{Z}) = -18.28 \pm 0.14$. The difference in log evidence between the two setups is thus $\Delta \log(\mathcal{Z}) = -0.34 \pm 0.16$. For
the log prior scans, the difference in log evidence between the two configurations is even lower, $\Delta \log(\mathcal{Z}) = -0.08 \pm 0.14$. As this
systematic difference is comparable with the statistical uncertainty, we can conclude that the posterior mass which is being missed by the standard,
Bayesian setup is negligible (especially so given that the systematic difference is much less than the typical difference in thresholds on the
Jeffreys' scale for the strength of evidence, which are at $\Delta \log(\mathcal{Z}) = 1.0, 2.5, 5.0$, see e.g.~\cite{Trotta:2008qt}). This leads us
to reject the speculation in Ref.~\cite{2010JHEP...04..057A} regarding the validity of posterior probability distributions obtained with the standard
configuration of {\sc MultiNest}.

The 2-D profile likelihood plots (bottom panels in Fig.~\ref{fig:cmssm_nlive_20k_tol_01}) should be compared with Fig.~1(a) in
Ref.~\cite{2010JHEP...04..057A}, which were obtained with the GA\footnote{We have verified that reducing  $n_{\rm live} = 4,000$,
while keeping $\mathrm{tol} = 1 \times 10^{-4}$, leads to profile likelihoods very similar to the ones shown in
Fig.~\ref{fig:cmssm_nlive_20k_tol_01}, albeit slightly more noisy. We thus do not show those results, but we remark that this configuration reduces the computational effort by a factor of $\sim 3$ compared with the case where $n_{\rm live} = 20,000$, without degrading the profile likelihood results appreciably.}.  Our best-fit point
(i.e, the unconditional MLE) has $\chi^2 = 9.26$ and lies in the focus point region. Therefore,  1$\sigma$ and 2$\sigma$ regions
are approximately delimited by $\chi^2 < 11.56$ and $\chi^2 < 15.43$, respectively.    The {\sc MultiNest} best-fit point has a
slightly better $\chi^2$ value than the GA best-fit point ($\chi^2 = 9.35$). However, {\sc MultiNest} 1$\sigma$ and 2$\sigma$
regions are much larger than the corresponding GA regions. This is because {\sc MultiNest} has been able to map out more
accurately the regions surrounding the best-fit than GA. By comparing 2-D posterior and profile likelihood plots in
Fig.~\ref{fig:cmssm_nlive_20k_tol_01}, we notice that most of the 1$\sigma$ profile likelihood interval in the focus point region of the  $m_{1/2}-m_0$ lies outside of the corresponding 2$\sigma$ Bayesian credible interval. This is a clear sign that the high
likelihood region in the focus point region is spike-like. These regions are not well explored by {\sc MultiNest} with the
standard configuration ($\mathrm{tol} = 0.5$). If one is interested in mapping out the posterior, however, this is fully
justified, as these regions contribute almost nothing to the Bayesian evidence or the posterior probability. If however one
wants to use {\sc MultiNest} for profile likelihood intervals, then it is imperative to map out those regions in much greater
detail.

The values of global best-fit point and the best-fit point in stau co-annihilation region (COA) found by {\sc MultiNest} are listed in Tab.~\ref{tab:BFPval}
and have similar parameter values to ones found by GA. Fig.~\ref{fig:cmssm_profile_1D}, shows the 1-D profile likelihood. This figure should be compared with
the corresponding 1-D profile likelihoods obtained with the GA method in Fig. 3 of Ref.~\cite{2010JHEP...04..057A}.



\begin{figure}
\begin{center}
\subfigure[]{\includegraphics[width=0.8\columnwidth]{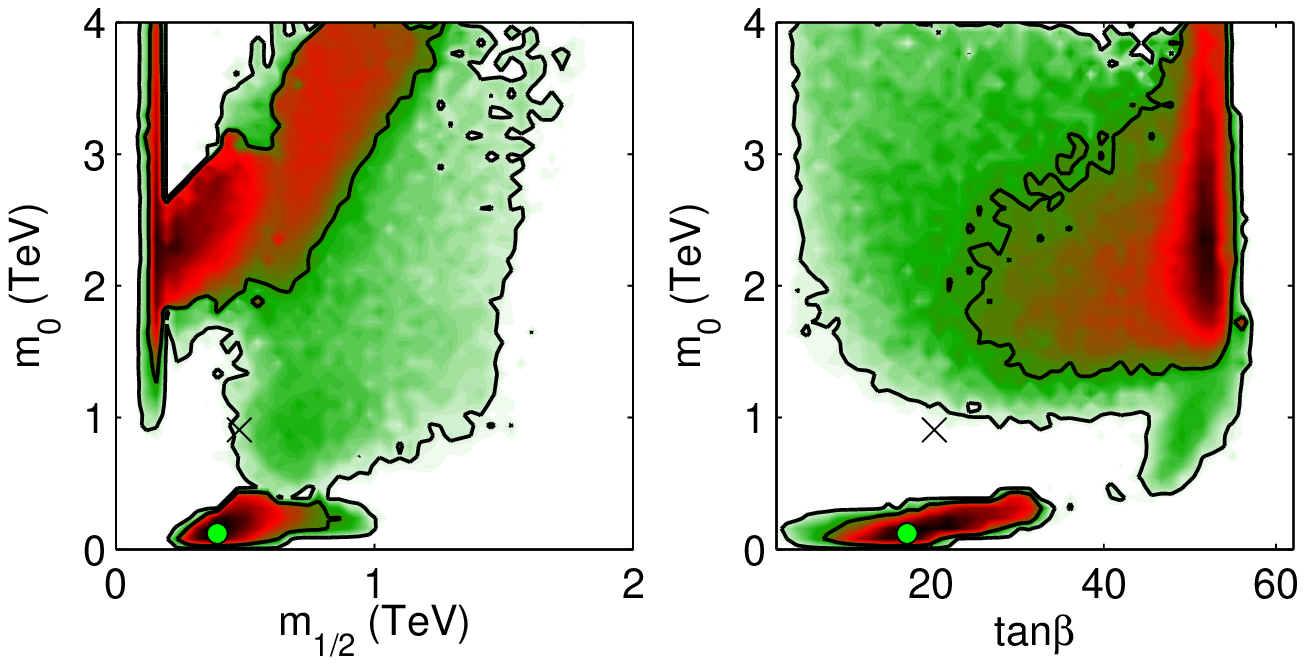}}\\
\subfigure[]{\includegraphics[width=0.8\columnwidth]{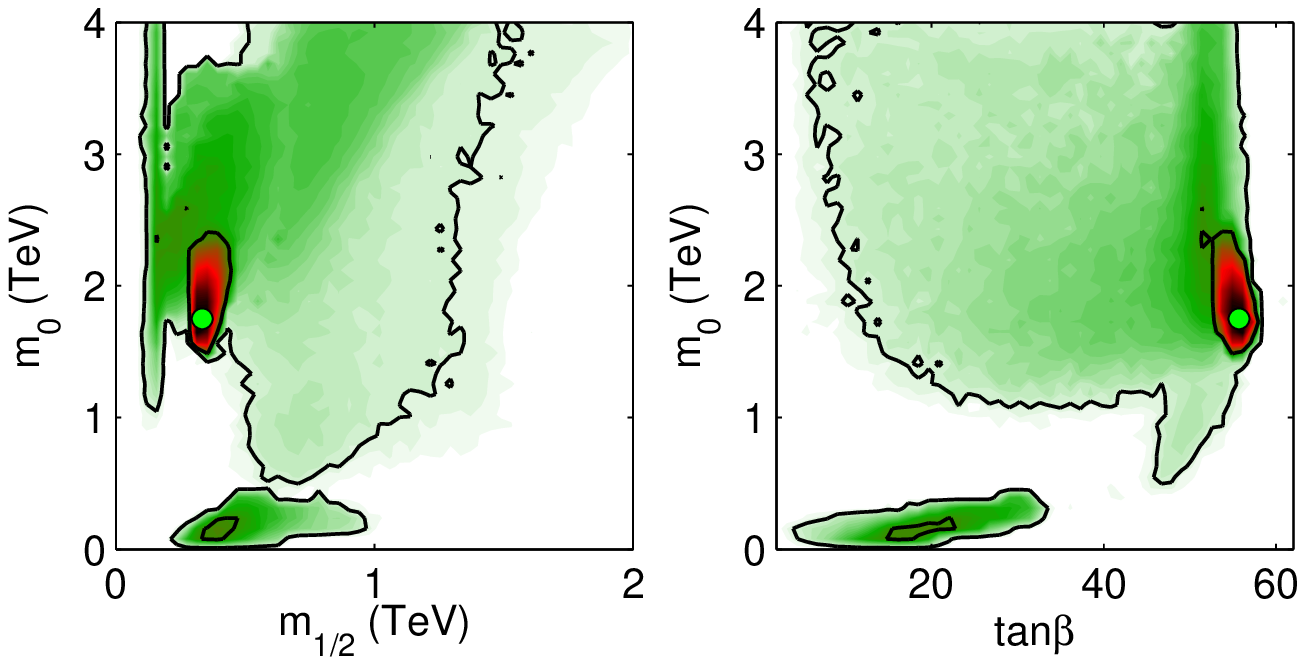}}
\caption{2-D profile likelihood functions for (a) log priors and (b)
  flat+log merged chains for CMSSM parameters. The contours represent
  the 68\% and 95\% profile likelihood confidence intervals. 20,000
  live points were used with $\mathrm{tol}$ set to $1 \times
  10^{-4}$. The mean and best-fit parameters values are shown by black
  cross and green circle respectively.}
\label{fig:cmssm_log_merged}
\end{center}
\end{figure}

\begin{figure}
\begin{center}
\includegraphics[width=1\columnwidth]{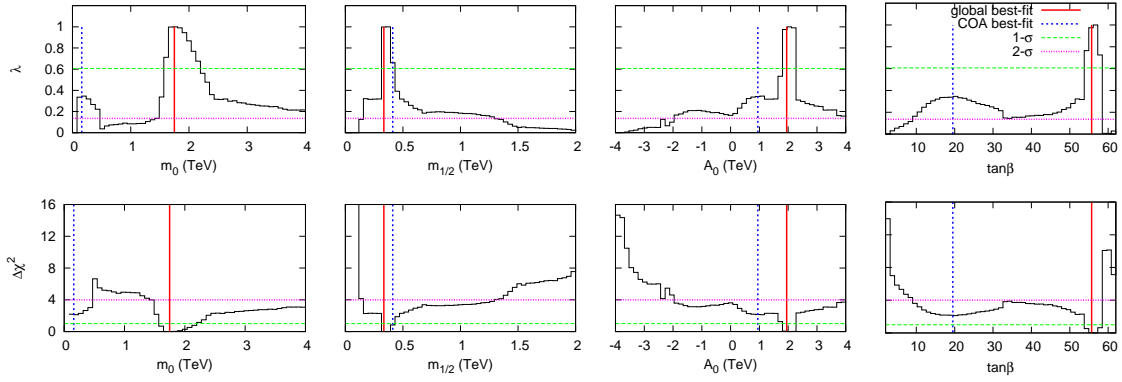}
\caption{1-D profile likelihoods for the CMSSM parameters for flat+log merged chains, normalized to the global best-fit point.
  The red solid and blue dotted vertical lines represent the global best-fit point ($\chi^2 = 9.26$, located in the focus point
  region) and the best-fit point found in the stau co-annihilation region ($\chi^2 = 11.38$) respectively. The upper and lower
  panel show the profile likelihood and $\Delta\chi^2$ values, respectively.  Green (magenta) horizontal lines represent the
  $1\sigma$ ($2\sigma$) approximate confidence intervals. {\sc MultiNest} was run with flat priors, 20,000 live points and
  $\mathrm{tol}=1 \times 10^{-4}$.}
\label{fig:cmssm_profile_merged_1D}
\end{center}
\end{figure}

\TABLE[t]{
\centering
{\footnotesize
\begin{tabular}{|l | l | l|} \hline
 & Global BFP & \multicolumn{1}{c|}{COA BFP} \\
 & (located in FP region) & \multicolumn{1}{c|}{} \\ \hline\hline
\multicolumn{3}{|c|}{Model and Nuisance Parameters} \\ \hline
$m_0$           &  $1748.7$~GeV    & $159.0$~GeV \\
$m_{1/2}$       &  $334.1$~GeV    & $411.4$~GeV \\
$A_0$           &  $1949.5$~GeV    & $946.2$~GeV \\
$\tan\beta$     &  $55.7$    & $19.53$ \\
$m_t$       &  $173.0$~GeV    & $173.3$~GeV \\
$m_b(m_b)^{\overline{MS}}$ & $4.19$~GeV  & $4.20$~GeV \\
$\alpha_s^{\overline{MS}}(M_Z)$       &  $0.1178$ & $0.1182$ \\
$1/\alpha^{\overline{MS}}(M_Z)$  & $127.952$ & $127.959$ \\ \hline
\multicolumn{3}{|c|}{Observables} \\ \hline
$m_W$     &  $80.367$~GeV   & $80.371$~GeV \\
$\sin^2 \theta_{\rm{eff}}$    &  $0.23156$      & $0.23153$ \\
$\delta a_{\mu}^{\text{SUSY}} \times 10^{10}$ & 6.8 & 13.7 \\
$BR(\overline{B}\rightarrow X_s\gamma) \times 10^{4}$ & 3.45 & 2.95 \\
$\Delta M_{B_s}$   &  $17.31~\mbox{ps}^{-1}$  & $19.0~\mbox{ps}^{-1}$ \\
$BR(\overline{B}_u\to \tau \nu) \times 10^{4}$ &  $1.38$  & $1.46$ \\
$\Omega_{\chi} h^2$ &  0.11408 & 0.11038 \\
$BR(\overline{B}_s\to\mu^+\mu^-)$ & $ 4.08\times 10^{-8}$ & $ 3.89\times 10^{-8}$\\ \hline
\end{tabular}
}
\caption[aa]{\footnotesize{Best-fit parameter and observable values found from flat+log merged chains.  These
quantities are shown for both the global best-fit point, located in the focus point (FP) region, as well as the best-fit point in the
stau co-annihilation (COA) region.}} \label{tab:BFPval}
}

In principle, the profile likelihood does not depend on the choice of priors. However,  in order to explore the parameter space
using any Monte Carlo technique, a set of priors needs to be defined. Any numerical sampling technique effectively defines a metric on the parameter space of interest, and different choices of this metric will generally lead to different regions of
the parameter space to be explored in greater or lesser detail. Therefore, different choice of priors might lead to slightly
different profile likelihoods.   Uniform priors in $\log(m_0)$ and $\log(m_{1/2})$ (called $\log$ priors) are often employed in
Bayesian analyses of CMSSM as the masses are scale parameters. In order to get a feeling for the robustness of our profile
likelihoods, we show the 2-D profile likelihoods from a scan using  $\log$ priors in Fig.~\ref{fig:cmssm_log_merged}(a). As
expected, the parameter space with lower $m_0$ and $m_{1/2}$ values has been explored more thoroughly. The best-fit point found
has $\chi^2 = 11.44$ and lies in the stau co-annihilation region (COA), thus this scan with $\log$ priors has missed a a higher likelihood point in the focus point region.  Conversely, the scan with flat priors has explored the COA region in similar detail.  Thus, while these tunings of {\sc MultiNest} are clearly superior for profile likelihood analysis than the default settings for Bayesian analysis, there remains sensitivity to the choice of priors used in the scans as previously seen in the literature. 

We can obtain more robust profile likelihoods by simply merging the flat and $\log$ prior scans, the performing the profiling over the joint set of
samples (encompassing about 10.85 million points).  This does not come at a greater computational cost given that a responsible Bayesian analyses
would estimate sensitivity to the choice of prior as well.  The resulting 2-D and 1-D profile likelihoods are shown in
Figs.~\ref{fig:cmssm_log_merged}(b) and ~\ref{fig:cmssm_profile_merged_1D} respectively. The most obvious difference between the merged and flat
profile likelihood distributions is the appearance of the funnel region at low $m_{1/2}$ values which was not explored thoroughly with flat priors
because of the its smaller probability mass. Apart from this, it is clear that the merged profile likelihood distributions are quite similar to the
distributions obtained with flat prior (see Fig.~\ref{fig:cmssm_nlive_20k_tol_01}(b)).

\begin{figure}
\begin{center}
\includegraphics[width=1\columnwidth]{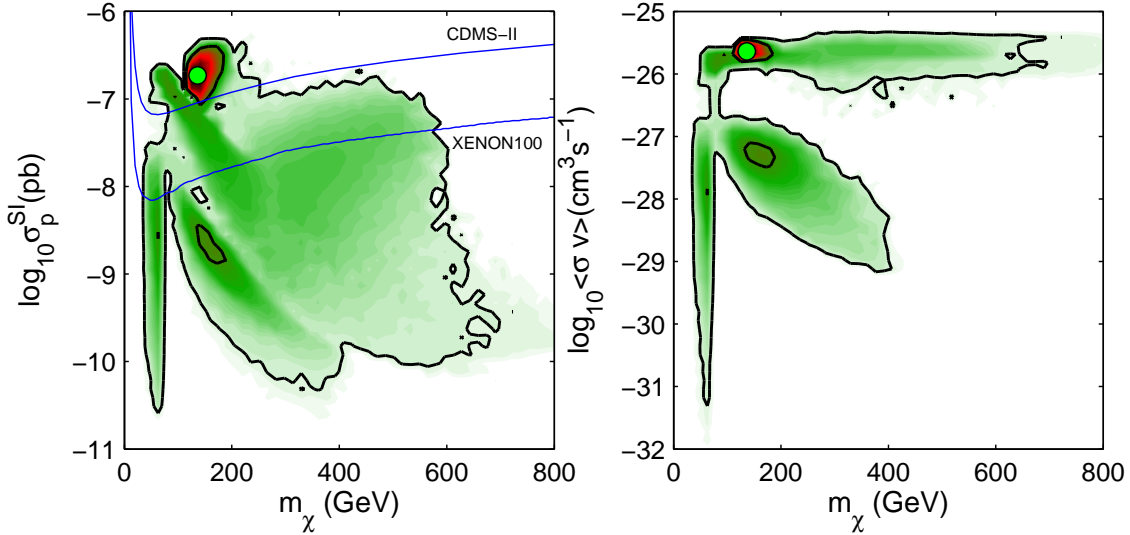}
\caption{2-D profile likelihood functions for flat+log merged chains for spin-independent scattering cross-section of the neutralino and a proton
$\sigma_{\rm p}^{\rm SI}$ versus the neutralino mass $m_{\chi}$ (left panel) and velocity-averaged neutralino self-annihilation cross-section $\left<\sigma
v\right>$ versus the neutralino mass $m_{\chi}$ (right panel). The contours represent the 68\% and 95\% profile likelihood confidence intervals. 20,000 live points
were used with $\mathrm{tol}$ set to $1 \times 10^{-4}$. The best-fit parameters values are shown by green circle. The latest experimental exclusion
limits at the 90\% confidence level from CDMS-II \cite{2009PhRvL.102a1301A} and XENON100 \cite{2011arXiv1104.2549X} are also shown in the left panel.}
\label{fig:cmssm_profile_DD_2D}
\end{center}
\end{figure}

In the left panel of Fig.~\ref{fig:cmssm_profile_DD_2D}, we show the 2-D profile likelihood functions spin-independent scattering cross-section of the neutralino and
a proton $\sigma_{\rm p}^{\rm SI}$ versus the neutralino mass $m_{\chi}$ along with the latest experimental exclusion limits at the 90\% confidence
level from  CDMS-II \cite{2009PhRvL.102a1301A} and XENON100 \cite{2011arXiv1104.2549X}, derived under standard halo assumptions (and in particular, assuming a local DM density $\rho_\chi = 0.3$ GeV/cm$^3$). Our global best-fit point has
quite a large cross-section ($1.85 \times 10^{-7}$ pb) and almost the entire $1\sigma$ region in the FP region would already be excluded by current experimental
limits if one takes them at face value. However, there are several uncertainties that need to be accounted for in order to achieve a robust exclusion: astrophysical uncertainties in the local DM density and velocity distribution~\cite{Strigari:2009zb,McCabe:2010zh,Green:2010gw}, systematic uncertainties related to our position in the Milky Way halo~\cite{Pato:2010yq}, clumpiness of the DM distribution and impact of baryonic physics~\cite{Ling:2009eh}, hadronic matrix elements uncertainties~\cite{Ellis:2008hf}. Taken together, all those uncertainties can easily exceed an order of magnitude. For this reason, we have chosen not to include the current direct detection limits in the likelihood function, although we notice qualitatively that the FP best-fit point is coming under pressure from such limits. On the other hand, the best-fit $\sigma_{\rm p}^{\rm SI}$ in the COA region is relatively small ($2.02 \times 10^{-9}$~pb) and is well below the current
experimental limits.

Regarding indirect detection prospects, we plot the 2-D profile likelihood functions for the velocity-averaged neutralino self-annihilation
cross-section $\left<\sigma v\right>$ against the neutralino mass $m_{\chi}$ in the right panel of Fig.~\ref{fig:cmssm_profile_DD_2D}. Although there is a small
region of the parameter space around 300 GeV $\le m_{\chi} \le$ 500 GeV and $-27.5 < \log_{10} \left<\sigma v\right> < -26.5$ with high likelihood
points found by GA in \cite{2010JHEP...04..057A} and missed by {\sc MultiNest}, in general we have mapped the profile likelihood more thoroughly as
can be seen by comparing the right panel of Fig.~\ref{fig:cmssm_profile_DD_2D} with Fig.~6(a) in \cite{2010JHEP...04..057A}. Our global best-fit point has
$\left<\sigma v\right> = 2.29 \times 10^{-26}$cm$^3$ s$^{-1}$ which is very similar to the global best-fit point found in \cite{2010JHEP...04..057A}
and therefore we reach the same conclusion, namely that it might be possible to cover part of this high-likelihood FP region in the future with the
Large Area Telescope (LAT)\cite{2009ApJ...697.1071A} aboard the Fermi gamma-ray space telescope.

\TABLE[t]{
\centering
{\footnotesize
\begin{tabular}{|l | l | l|} \hline
 & Global BFP & \multicolumn{1}{c|}{COA BFP} \\
 & (located in FP region) & \multicolumn{1}{c|}{} \\ \hline\hline
$m_{\chi}$           &  $136.34$~GeV    & $156.46$~GeV \\ \hline
\multicolumn{3}{|c|}{Direct Detection} \\ \hline
$\sigma_{\rm p}^{SI}$           &  $1.85 \times 10^{-7}$~pb    & $2.02 \times 10^{-9}$~pb \\
$\sigma_{\rm p}^{SD}$       &  $1.96 \times 10^{-6}$~pb    & $4.05 \times 10^{-6}$~pb \\
$\sigma_{\rm n}^{SD}$           &  $1.34 \times 10^{-6}$~pb    & $3.00 \times 10^{-6}$~pb \\ \hline
\multicolumn{3}{|c|}{Indirect Detection} \\ \hline
$\left<\sigma v\right>$     &  $2.29 \times 10^{-26}$ cm$^3$ s$^{-1}$   & $4.89 \times 10^{-28}$ cm$^3$ s$^{-1}$ \\ \hline
\end{tabular}
}
\caption[aa]{\footnotesize{Best-fit neutralino mass and dark matter direct and indirect detection observables. These quantities are
shown for both the global best-fit point, located in the focus point (FP) region, as well as the best-fit point in the stau co-annihilation (COA)
region.}} \label{tab:DDBFPval}
}


\section{Conclusions}\label{sec:conclusions}

As the LHC impinges on the most anticipated regions of SUSY parameter space, 
the need for statistical techniques that will be able to cope with the complexity of SUSY phenomenology
is greater than ever.  Early on, the data will not be strong enough to dominate the impact of priors on
Bayesian inference; thus, the field is preparing to present the complementary information provided by
Bayesian techniques and the traditional results from the profile likelihood.


In this paper we have shown that, when configured appropriately,  {\sc
  MultiNest} can be succesfully employed for approximating the profile likelihood functions, even
though it was primarily designed for Bayesian analyses.  
  In particular, we have demonstrated that it is important to use a
termination criterion that allows {\sc MultiNest} to explore
high-likelihood regions. We have also shown that the likelihood spikes which play a key role
in the profile likelihood and are missed by {\sc MultiNest} when it is run
in Bayesian analysis mode, have no bearing on the posterior
distribution.  Despite these significant improvements, we do find that
the prior used in the  {\sc MultiNest} scan can influence the profile likelihood ratio, 
particularly if the global maximum of the likelihood is spiky and in a region with extremely low prior probability.
This can be partially abated by merging samples from scans with different priors, which are often
available from studies of prior dependence in the corresponding Bayesian results.

We compared our results and conclusions to those reported in  Ref.~\cite{2010JHEP...04..057A}, which used genetic algorithms  for mapping the profile
likelihood functions.  The authors of that study concluded that their sampling algorithm resulted in better approximations to the profile likelihood
than those obtained with samples from Bayesian techniques like MCMC and {\sc MultiNest}.  They also speculated that {\sc MultiNest} might not  be
sufficiently accurate even for Bayesian analyses. Our results indicate that when run appropriately, {\sc MultiNest} provides a significantly better
approximation to the profile likelihood than the genetic algorithm method presented in Ref.~\cite{2010JHEP...04..057A}, particularly in the
exploration of the parameter space near the boundaries of the interval, albeit at a slightly increased computational cost. There are also indications that  {\sc MultiNest} might be more reliable in identifying the overall best-fit point than the genetic algorithm, as the latter only found the overall best-fit in 1 out of 10 runs (see Fig.~8 in Ref.~\cite{2010JHEP...04..057A}).   


We end by speculating that the ideal algorithm for profile likelihood based analyses in complex, multi-modal problems such as supersymmetric parameter spaces may require a hybrid
of global scans like MCMC and  {\sc MultiNest} together with dedicated maximization packages like  {\sc Minuit} \cite{James:1975dr}.  The challenge for most
maximization packages -- whether they are based on gradient descent,  conjugate gradient descent,  or the EM-algorithm --  is that they can get stuck in
local maxima.  Hence, it seems clear that a more global scan will be required.  However, once the scanning algorithm has located a basin of attraction the
maximization packages are likely to be the best tool for refining the  local maxima.  As we discussed here,  {\sc MultiNest} with a low tolerance acts as a
maximizer, and can be subject to similar problems with local maxima. The nested structure provided by the nested sampling algorithm may provide an efficient
and practical way for defining those basins of attraction, reducing the number of points that will initiate a dedicated maximization.  In future work we wish
to investigate these hybrid approaches and compare {\sc Minuit} and {\sc MultiNest} for this dedicated maximization step.

\section*{Acknowledgements}
We thank the organizers of the PROSPECTS workshop (Stockholm, Sept
2010) for a stimulating meeting that offered the opportunity to
discuss several of the issues presented in this paper, and Pat Scott for useful comments on an earlier draft.  This work was
carried out largely on the {\sc Cosmos} UK National Cosmology
Supercomputer at DAMTP, Cambridge and the Darwin Supercomputer of the
University of Cambridge High Performance Computing Service ({\tt
  http://www.hpc.cam.ac.uk/}), provided by Dell Inc. using Strategic
Research Infrastructure Funding from the Higher Education Funding
Council for England. FF is supported by a Research Fellowship from
Trinity Hall, Cambridge.  K.C. is supported  by the US National Science Foundation grants PHY-0854724 and PHY-0955626.

\providecommand{\href}[2]{#2}\begingroup\raggedright\endgroup

\label{lastpage}

\end{document}